\documentclass[12pt,a4paper]{article}
\usepackage{graphicx}

\begin{document}
\vspace {10 mm}


\begin{center}
{ \bf {\large On the double and triple-humped fission barriers \\  
and half-lives of actinide elements}}

\bigskip
{\large C. Bonilla, G. Royer}

\end{center}

\begin{abstract}
The deformation barriers standing in the quasi-molecular shape path have been 
determined in the actinide region within a macroscopic-microscopic energy
 derived from a generalized liquid drop model,
 the algebraic droplet model shell corrections and analytic expressions for the pairing energies. 
Double and triple-humped fission barriers appear. The second barrier 
corresponds to the transition from one-body shapes to two touching ellipsoids. 
The third minimum and third peak, when they exist, come from shell rearrangements in the deformed fragment.
 The shape of the other almost magic one is close to the sphere. The barrier heights agree
with the experimental results, the energy of the second minimum being a little too high. 
The predicted half-lives follow the experimental data trend.  
\end{abstract}

\section{Introduction}
The possibility of transmutation of nuclear waste and of production of energy 
by accelerator-driven systems is under consideration. The knowledge of all the nuclear reactions which
constitute a non negligible part of the reaction cross section is needed \cite{duij01}. 
Different codes are under construction
or improvement (Fluka, Gnash, Talys,..) and accurate potential barriers must be calculated rapidly, particularly
in the actinide region, to predict or firstly to reproduce the mass and charge distributions governing
the fission cross sections \cite{hamb03}. Furthermore, new precise measurements renew also interest in
investigating the multiple-humped barriers of the actinide nuclei and heaviest elements. 
The analysis of the fission probability and of the angular distribution of the fission fragments
support the presence of hyperdeformed states in a deep third
well in several Th and U isotopes \cite{krasz98,krasz99,huny01} confirming the pioneering work of Blons et al
 \cite{blon84} in $^{231,233}$Th.  The observed strongly enhanced low energy $\alpha$ decay in some 
heavy actinide nuclei is also
explained by transition from a third hyperdeformed minimum and the possibility that the third minimum is the 
true ground state of very heavy and perhaps superheavy nuclei is even also advocated \cite{mari03}. In medium mass
nuclei some signs of hyperdeformed rotational bands have been found, but no discrete HD level has been identified 
 \cite{gali93,roye93}.

By adding at the macroscopic liquid drop model energy of elongated one-body shapes an oscillatory microscopic 
contribution, the Strutinsky's method \cite{stru67} generated 
double-humped barriers allowing to predict and explain the fission isomer characteristics. 
Myers and Swiatecki \cite{myer67} proposed analytic formulae in the same objective
of taking into account the shell and pairing energies. Later on, the asymmetric two-center shell model \cite{must73}, 
Hartree-Fock-Bogoliubov \cite{berg81} 
and relativistic mean field theories \cite{blum94} have also allowed to reproduce double-humped barriers.
 The heights of the inner and asymmetric outer fission barriers are almost constant 
(5-6 MeV) from Th to Am isotopes \cite{bjor80,wage91}. It is a severe test for the theoretical models.
In the actinide region the third hyperdeformed minimum was 
predicted by M\"oller et al \cite{moll72} and by recent theoretical approaches \cite{beng87,bhan89,cwio94}. 

It has been previously shown within a Generalized Liquid Drop Model taking into account both the proximity energy 
between close opposite surfaces, the asymmetry and an accurate radius that  
most of the symmetric and asymmetric fission \cite{roye84,roye02}, $\alpha$ and light nucleus emissions 
\cite{roye00,roye01} and super and highly deformed state data \cite{roye03} can also be reproduced
 in the fusiolike shape path. 
The purpose 
of this work is, within this GLDM, to go beyond the two
separated sphere approximation by taking into account the ellipsoidal deformations 
of the two different fission 
fragments and their associated shell and pairing energies, investigating all the possible mass and charge asymmetries. 

\section{Potential energy}
The total energy of a deformed nucleus is the sum of the
  macroscopic GLDM energy, the shell correction energy and the pairing energy.
The GLDM energy is given by \cite{roye85}
\begin{equation}
E=E_V+E_S+E_C+E_{prox}+E_{rot},
\end{equation}  
where the different terms are respectively the volume, surface, Coulomb, nuclear proximity and rotational
energies.

All along the deformation path the nuclear proximity energy term $E_{prox}$
allows to take into account the effects of the attractive nuclear forces
between nucleons facing each other across a neck in the case of a deformed one-body shape or across a gap
in the case of two separated fragments.  
This is not a small correction in the quasi-molecular shape path. 
For example, at the contact point between two
spherical Kr and Ba nuclei the proximity energy reaches $-43$ MeV.
\begin{equation}
E_{prox}(r)=2\gamma \int \Phi \left \lbrack D(r,h)/b\right 
\rbrack 2 \pi hdh.
\end{equation}  
$r$ is the distance between the mass centres.
$h$ is the transverse distance varying from the neck radius
to the height of the neck border. $D$ is the distance between the
opposite surfaces and $b$ the surface width. $\Phi$ is the proximity function
and $\gamma$ the surface parameter. 

\subsection{One-body-shapes}
For one-body shapes, the first three contributions are given by 
\begin{equation}                                                             
E_V=-15.494(1-1.8I^2)A \ MeV,
\end{equation}
\begin{equation}
E_S=17.9439(1-2.6I^2)A^{2/3}\frac{S}{4\pi R^2_0}\ MeV,
\end{equation}
\begin{equation}
E_C=0.6e^2(Z^2/R_0)B_C.
\end{equation}
 $B_C$ is the Coulomb shape dependent function, S is the surface and
 $I$ is the relative neutron excess \cite{roye85}. \newline
\begin{equation}
B_C=0.5\int (V(\theta)/V_0)(R(\theta)/R_0)^3\sin \theta d \theta,
\end{equation}  
 where $V(\theta )$ is the electrostatic potential
 at the surface and $V_0$ the surface potential of the sphere.\newline
The radius $R_{0}$ of the compound nucleus is defined as:
\begin{equation}
R_0=(1.28A^{1/3}-0.76+0.8A^{-1/3}) \  fm.
\end{equation}  
which leads, for example, to
 $R_0=5.3$ fm and $r_0=1.15$ fm for $^{98}$Zr and 
 $R_0=7.5$ fm and $r_0=1.18$ fm for $^{255}$Fm.
The radius of the two fragments is calculated assuming volume conservation.    

\begin{figure}[ht]
\begin{center}
\includegraphics[width=8cm]{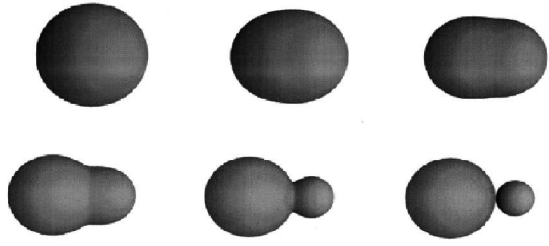}
\caption{Selected shape sequence to simulate the one-body shape evolution.}
\label{lemniscatoids}
\end{center}
\end{figure}

As in previous works, the one-body shape sequence is described within two joined elliptic 
lemniscatoids which allow to simulate the development of a deep neck in compact and little elongated
 shapes with spherical ends (see Fig. \ref{lemniscatoids}). The proximity energy is maximized in this deformation path.

\subsection{Two separated ellipsoids}
\begin{figure}[ht]
\begin{center}
\includegraphics[width=8cm]{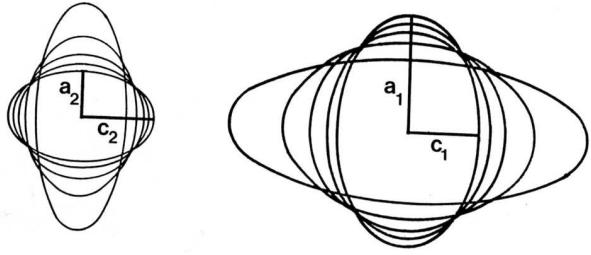}
\caption{Two coaxial ellipsoid configuration describing the two-body shape part of the fission barrier.
 The fission axis is the common axis of revolution.}
\label{ellipsoids}
\end{center}
\end{figure}

For two-body shapes, the coaxial ellipsoidal deformations   
have been considered \cite{roye92} (see Fig. \ref{ellipsoids}). 
The system configuration depends
on two parameters : the ratios $s_i$ ($i=1,2$) between the transverse semi-axis $a_i$ and the radial semi-axis $c_i$
of the two different fragments.
\begin{equation}
a_i=R_is^{1/3}_i \ \ and \ \   c_i=R_is^{-2/3}_i.  
\end{equation}

The prolate deformation is  characterized by $s\le1$ and the related eccentricity is written as $e^2=1-s^2$ while in 
the oblate case $s\ge1$ and $e^2=1-s^{-2}$.\newline
The volume and surface energies are $E_{V_{12}}=E_{V_1}+E_{V_2}$ and $E_{S_{12}}=E_{S_1}+E_{S_2}$.\newline  
In the prolate case, the relative surface energy reads 
\begin{equation}
B_{Si}=\frac{(1-e^2_i)^{1/3}}{2}\left\lbrack 1+\frac{sin^{-1}(e_i)}{e_i(1-e^2_i)^{1/2}}\right \rbrack
\end{equation} 
and in the oblate case
\begin{equation}
B_{Si}=\frac{(1+{\epsilon}^2_i)^{1/3}}{2}\left\lbrack 1+\frac{ln({\epsilon}_i+(1+{\epsilon}^2_i)^{1/2})}
{{\epsilon}_i(1+{\epsilon}^2_i)^{1/2}}\right \rbrack \ \ \ \ {\epsilon}^2_i=s^2_i-1.
\end{equation}  
The Coulomb self-energy of the spheroid $i$ is
\begin{equation}
E_{C,self}=\frac{3e^2Z^2_iB_{ci}}{5R_i}.
\end{equation}
The relative self-energy is, in the prolate case 
\begin{equation}
B_{Ci}=\frac{(1-e^2_i)^{1/3}}{2e_i}ln\frac{1+e_i}{1-e_i}
\end{equation} 
and, in the oblate case
\begin{equation}
B_{Ci}=\frac{(1+{\epsilon}^2_i)^{1/3}}{{\epsilon}_i}tan^{-1}{{\epsilon}_i}.
\end{equation}  
The Coulomb interaction energy between the two fragments reads 
\begin{equation}
E_{C,int}=\frac{e^2Z_1Z_2}{r}\left\lbrack s({\lambda}_1)+s({\lambda}_2)-1+S({\lambda}_1,{\lambda}_2)\right\rbrack
\ \ \ \ {\lambda}^2_i=\frac{c^2_i-a^2_i}{r^2},
\end{equation}  
r being, as before, the distance between the two mass centres.\newline
In the prolate case, $s({\lambda}_i)$ is expressed as
\begin{equation}
s({\lambda}_i)=\frac{3}{4}(\frac{1}{{\lambda}_i}-\frac{1}{{\lambda}^3_i})ln
(\frac{1+{\lambda}_i}{1-{\lambda}_i})+\frac{3}{2{\lambda^2_i}},
\end{equation}  
while, for the oblate shapes, 
\begin{equation}
s({\lambda}_i)=\frac{3}{2}(\frac{1}{{\omega}_i}+\frac{1}{{\omega}^3_i})tan^{-1}
{\omega}_i-\frac{3}{2{\omega^2_i}} \ \ \ \ \ \ {\omega}^2_i=-{\lambda}^2_i.
\end{equation}  
$S({\lambda}_1,{\lambda}_2)$ can be represented in the form of a two-fold summation
\begin{equation}
S({\lambda}_1,{\lambda}_2)=\sum_{j=1}^{\infty}\sum_{k=1}^{\infty}\frac{3}{(2j+1)(2j+3)}\frac{3}{(2k+1)(2k+3)}
\frac{(2j+2k)!}{(2j)!(2k)!}{\lambda}^{2j}_1{\lambda}^{2k}_2.   
\end{equation}

\section{Shell energy}

The shell corrections for a deformed nucleus 
have been determined within the algebraic formulae given in the Droplet Model 
\cite{myer77} with slightly different values of the parameters.   
\begin{equation}
E_{shell}=E_{shell}^{sphere}(1-2.6\theta^{2})e^{-0.9\theta^{2}}.
\end{equation}
The shell corrections for a spherical nucleus are
\begin{equation}
E_{shell}^{sphere}=5.8\left \lbrack (F(N)+F(Z))/(0.5A)^{2/3}-0.26A^{1/3}\right 
\rbrack \  MeV,
\end{equation}
where, for $M_{i-1}<X<M_i$,
\begin{equation}
F(X)=q_i(X-M_{i-1})-0.6(X^{5/3}-M_{i-1}^{5/3}).
\end{equation}
$M_i$ are the magic numbers and
\begin{equation}
q_i=0.6(M_i^{5/3}-M_{i-1}^{5/3})/(M_i-M_{i-1}).
\end{equation}
The selected highest proton magic number is 114 while, for the two highest neutron magic numbers, the values 126 and 184
have been retained.
\begin{equation}
\theta^{2}=({\delta}R)^2/a^2.
\end{equation}
The distortion ${\theta}a$ is the root mean square of the deviation of the nuclear surface from the sphere, 
a quantity which incorporates indiscriminately all types of deformation. The range $a$ 
has been chosen to be $0.32r_0$.\newline
For the two-body shapes, the total shell energy is the sum of the shell corrections for each deformed fragment.

\section{Pairing energy}
The pairing energy has been calculated with the following expressions.\\
For odd Z, odd N and N=Z nuclei
\begin{equation}
E_{Pairing}=4.8/N^{1/3}+4.8/Z^{1/3}-6.6/A^{2/3}+30/A.
\end{equation}
For odd Z, odd N and $N \ne Z$ nuclei
\begin{equation}
E_{Pairing}=4.8/N^{1/3}+4.8/Z^{1/3}-6.6/A^{2/3}.
\end{equation}
For odd Z, even N nuclei
\begin{equation}
E_{Pairing}=4.8/Z^{1/3}.
\end{equation}
For even Z, odd N nuclei
\begin{equation}
E_{Pairing}=4.8/N^{1/3}.
\end{equation}
For even Z, even N nuclei
\begin{equation}
E_{Pairing}=0.
\end{equation}

\section{Potential barriers}
\begin{figure}[htbp]
\begin{center}
\includegraphics[height=10cm]{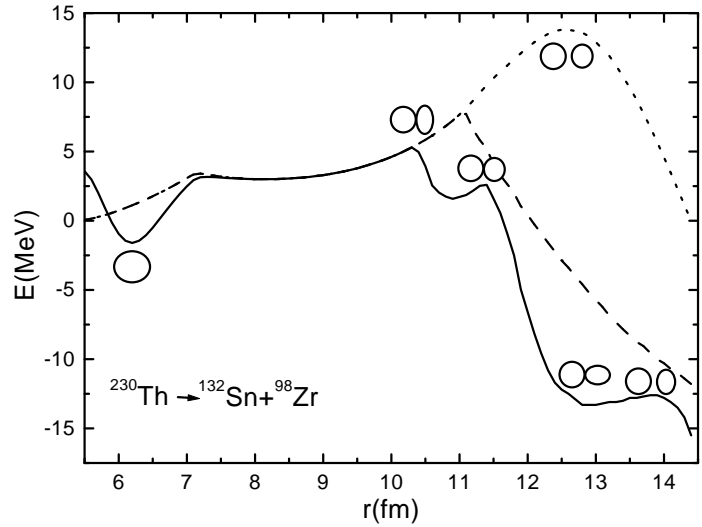}
\caption{Asymmetric fission barrier of a $^{230}$Th nucleus emitting a doubly magic nucleus $^{132}$Sn.
 The dashed and dashed-dotted curves give the energy within the two-sphere approximation for
the two-body shapes without and with shell corrections around the original sphere 
while the dotted and solid lines include the 
ellipsoidal deformations without and with shell energies. r is the distance between mass centres.}
\label{th230}
\end{center}
\end{figure}

The dependence of the potential barriers on the shape sequence and introduction of the microscopic
corrections is displayed in Fig. \ref{th230} for an asymmetric fission path of the $^{230}$Th nucleus.
The shell effects generate the slightly deformed ground state and contribute to the formation
of the first peak. 
The proximity energy flattens the potential energy curve and will explain
the formation of a deep second minimum for heaviest nuclei. 
The shell effects are attenuated at these large deformations.
In the exit channel corresponding
to the two-sphere approximation the top of the barrier ($r=12.6$ fm on this example) is reached after the rupture of
 the matter bridge between the two spherical fragments ($r=11.4$ fm). Then, the top corresponds to two
 separated spherical fragments maintained in unstable equilibrium by the balance 
between the attractive nuclear forces and the repulsive Coulomb ones. In this path, the introduction of
the shell and pairing effects for two-body shapes is not sufficient to reproduce accurately the experimental data
on the fission barrier heights of actinide nuclei.   
The transition between one-body and two-body shapes is less smooth when the ellipsoidal deformations 
of the fragments and the proximity energy are taken into account.
It corresponds to the passage (at $r=11$ fm for $^{230}$Th) from a one-body shape with spherical ends and a deep neck
 to two touching ellipsoidal fragments, one or both of them being slightly oblate. The barrier height is
 reduced by several MeV. The introduction of the shell effects still lowers the second peak and shifts it
to an inner position ($r=10.3$ fm here). It even leads to a
third minimum and third peak in this asymmetric decay path. A plateau appears also at larger distances around 10 MeV
below the ground state. 
It is due to the end of the contact between the two fragments and the persistence of the prolate
deformation of the lightest fragment.
 The end of the plateau corresponds to the rapid transition from prolate
to oblate shapes for the non-magical fragment and the vanishing of the proximity energy. This second 
 fragment returns to a prolate shape when the interaction Coulomb energy is smaller. 

\begin{figure}[htbp]
\begin{center}
\begin{tabular}{cc}
\includegraphics[height=5.5cm]{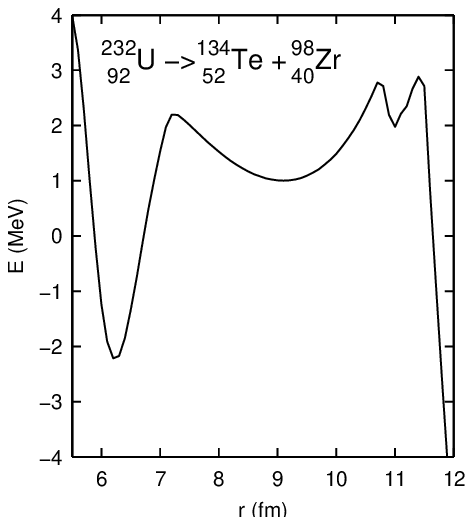} &
\includegraphics[height=5.5cm]{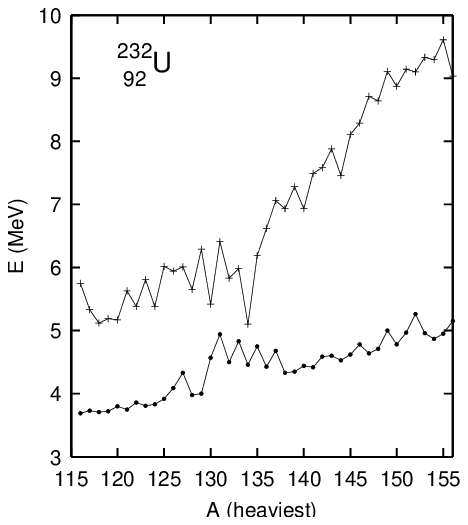} \\
\includegraphics[height=5.5cm]{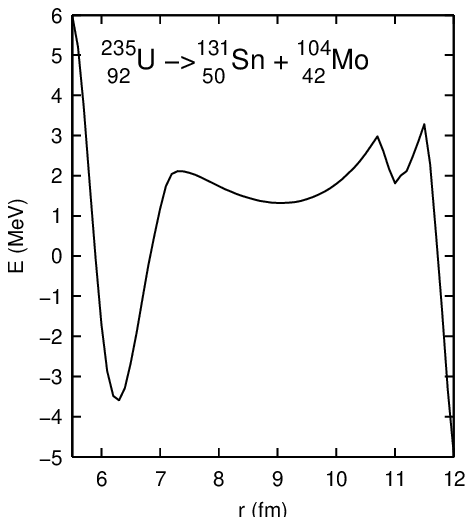} &
\includegraphics[height=5.4cm]{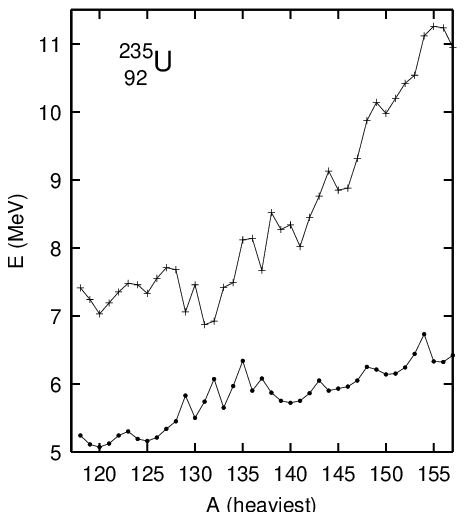} \\
\includegraphics[height=5.5cm]{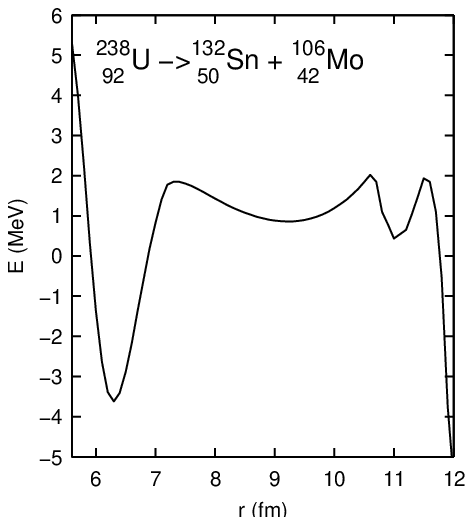} &
\includegraphics[height=5.4cm]{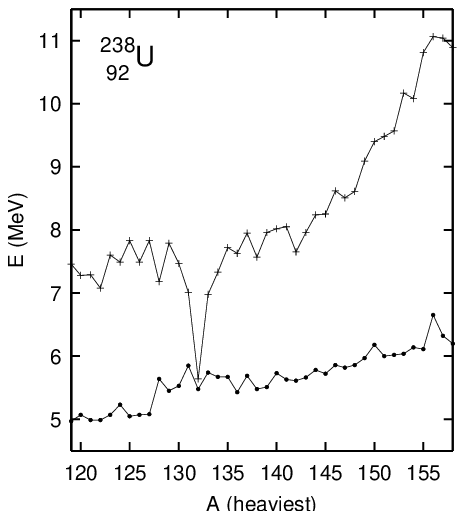}
\end{tabular}
\caption{On the left, multiple-humped fission barriers in the mentioned asymmetric fission path for $^{232,235,238}$U.
On the right, inner (full circles) and outer (crosses) fission barrier heights as a function of the mass 
of the heaviest fragment.}
\label{u232235238}
\end{center}
\end{figure}

\begin{figure}[htbp]
\begin{center}
\begin{tabular}{cc}
\includegraphics[height=5.5cm]{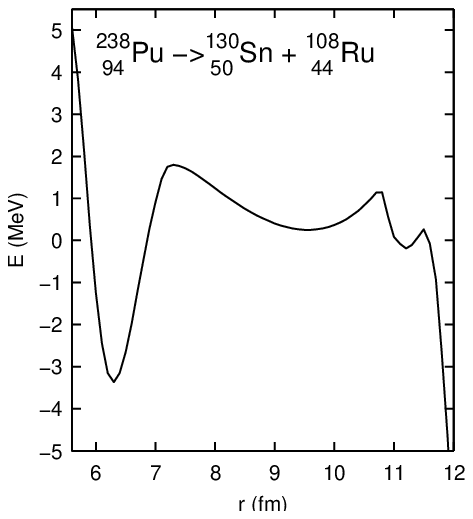} &
\includegraphics[height=5.5cm]{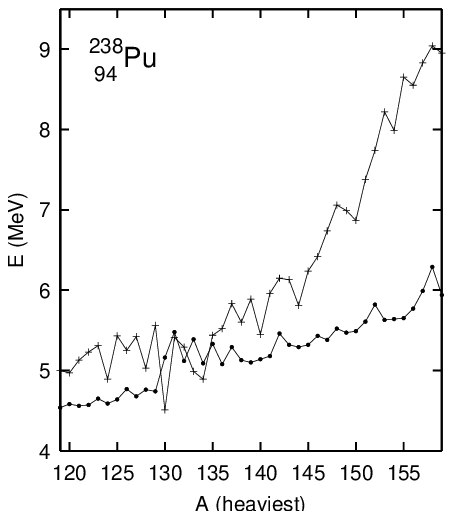} \\
\includegraphics[height=5.5cm]{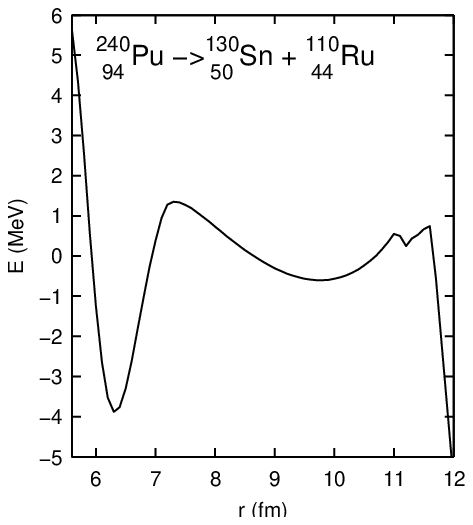} &
\includegraphics[height=5.5cm]{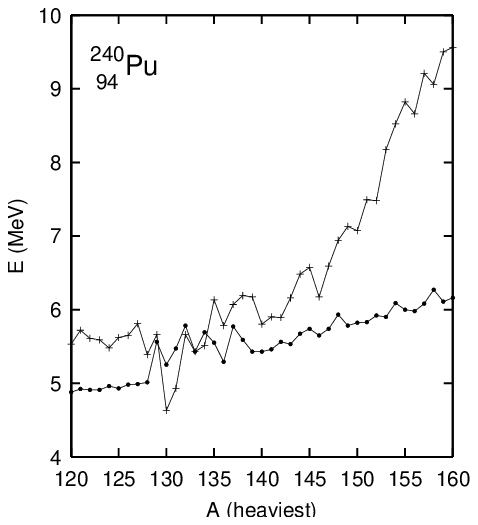} \\
\includegraphics[height=5.5cm]{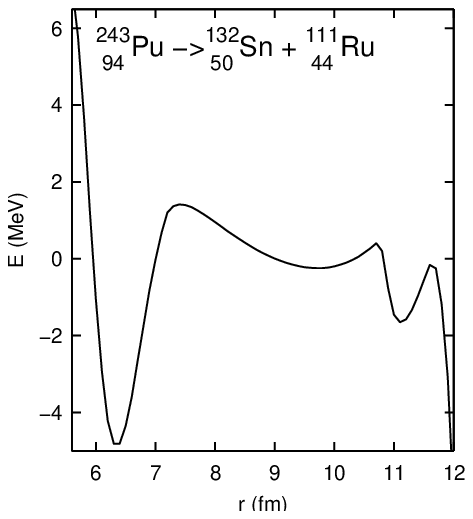} &
\includegraphics[height=5.5cm]{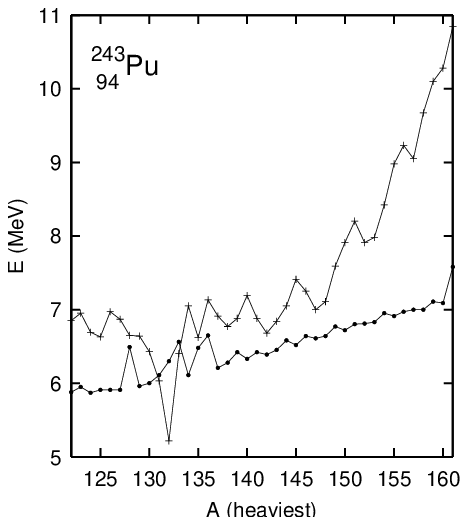}
\end{tabular}
\caption{On the left, multiple-humped fission barriers in the mentioned asymmetric fission path for $^{238,240,243}$Pu.
On the right, inner (full circles) and outer (crosses) fission barrier heights as a function of the mass 
of the heaviest fragment.}
\label{pu238240243}
\end{center}
\end{figure}

\begin{figure}[htbp]
\begin{center}
\begin{tabular}{cc}
\includegraphics[height=5.4cm]{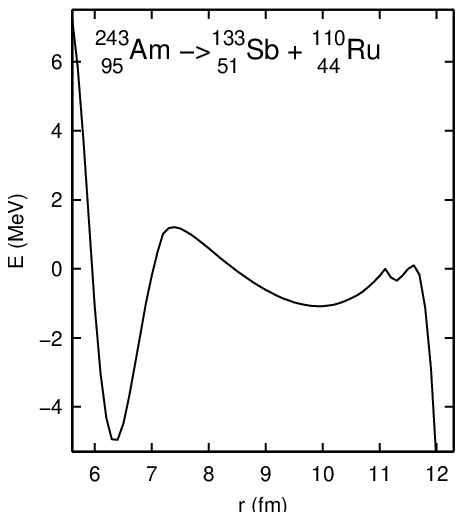} &
\includegraphics[height=5.5cm]{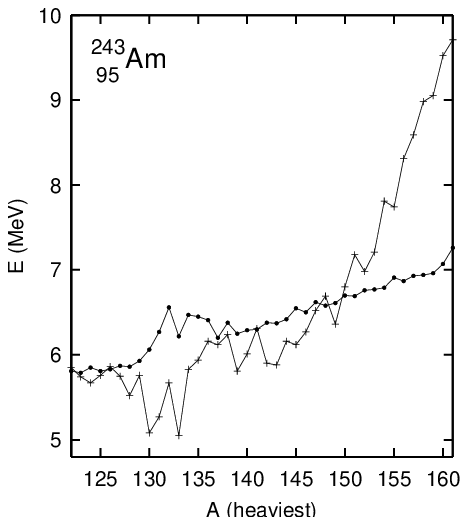} \\
\includegraphics[height=5.4cm]{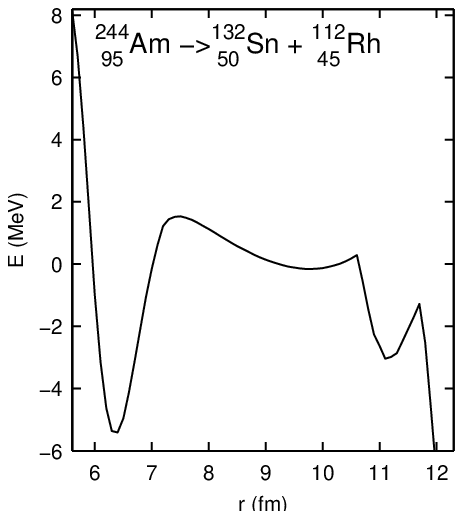} &
\includegraphics[height=5.5cm]{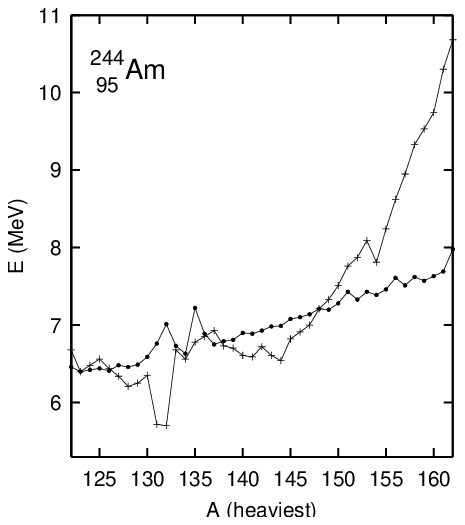} \\
\includegraphics[height=5.5cm]{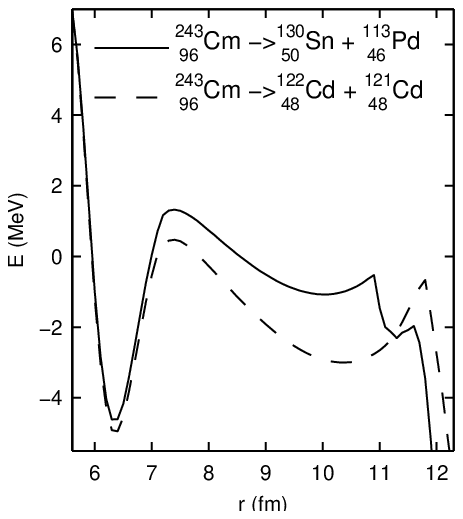} &
\includegraphics[height=5.5cm]{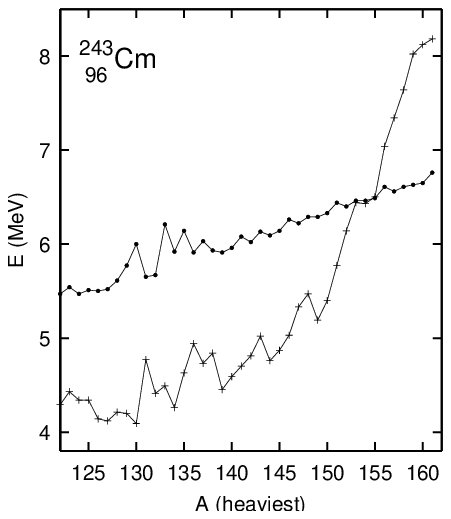}
\end{tabular}
\caption{Fission barriers for $^{243,244}$Am and $^{243}$Cm.
The inner and outer fission barrier heights are given, on the right, respectively by the full circles
and crosses.}
\label{am243244cm243}
\end{center}
\end{figure}

\begin{figure}[htbp]
\begin{center}
\begin{tabular}{cc}
\includegraphics[height=5.5cm]{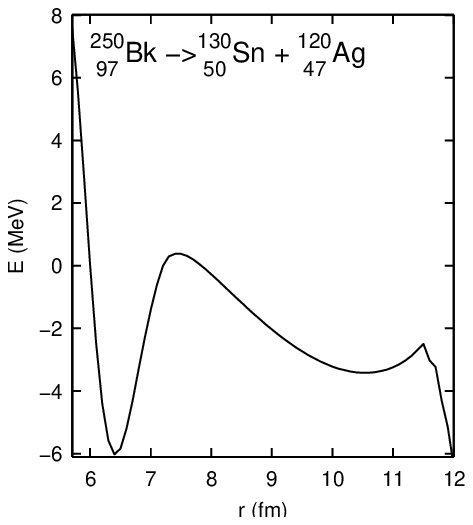} &
\includegraphics[height=5.5cm]{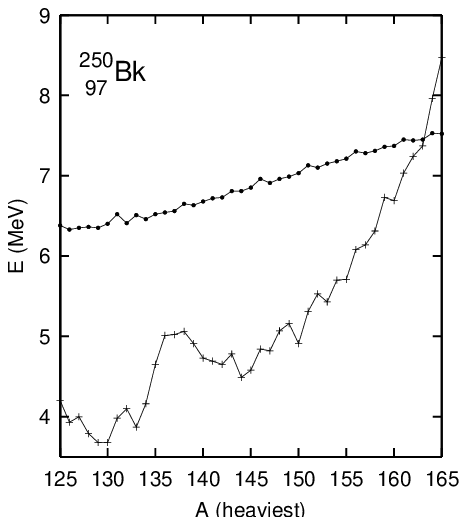} \\
\includegraphics[height=5.5cm]{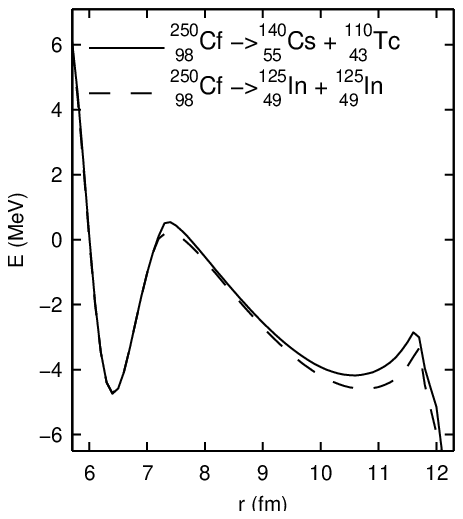} &
\includegraphics[height=5.5cm]{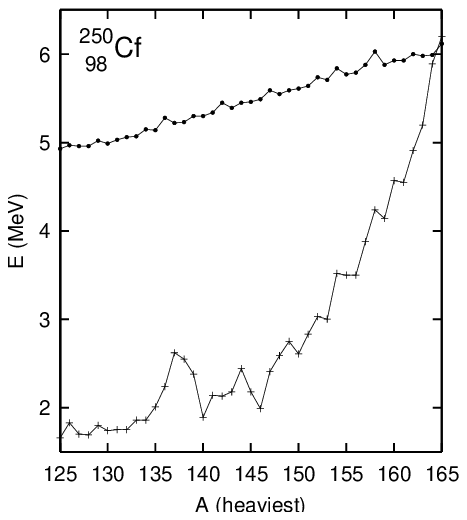} \\
\end{tabular}
\caption{Fission barriers for $^{250}$Bk and $^{250}$Cf.
The inner and outer fission barrier heights are given, on the right, respectively by the full circles
and crosses.}
\label{cf250bk250}
\end{center}
\end{figure}

The potential barriers for the $^{232,235,238}$U, $^{238,240,243}$Pu, $^{243,244}$Am, $^{243}$Cm, 
$^{250}$Bk and $^{250}$Cf nuclei are shown in Fig. \ref{u232235238} to \ref{cf250bk250}. 
It is important to mention that to obtain these barriers the only input parameters are $A_1, Z_1, A_2$
and $Z_2$ and that the calculations are very rapid on an usual computer and, consequently, can be integrated
in a more general and complex code.

For a given mass asymmetry, the charge asymmetry which minimizes the deformation energy has been selected. 
The proximity energy and the attenuated microscopic effects are responsible for the formation
 of a second one-body shape minimum. 
The heights of the two peaks generally increase with the asymmetry but the shell and pairing corrections induce strong 
variations from this global behaviour. Their main effect is to favour, for the U, Pu, Am and Cm isotopes, an asymmetric 
path where one fragment is close to the doubly magic number $^{132}_{50}$Sn nucleus, and, consequently, keeps an almost
 spherical shape. This effect is less pronounced for $^{243}$Cm and $^{250}$Cf since for nuclei with $Z\sim100$ the 
symmetric fission gives fragments with a charge of around 50. 
For these nuclei a symmetric and an asymmetric exit
channels are compared on the lower part of the Figures \ref{am243244cm243} and \ref{cf250bk250}.    
A third minimum and third peak appear in the asymmetric decay path. There is no third barrier
in the symmetric deformation path. The origin of the third peak is investigated in section VI.  

The calculated and experimental energies of the maxima and minima of the fission barriers are compared in table 1.
The choice of the most probable fission path is difficult for some isotopes since there is a true degenerescence
in energy between several paths of the multi-dimensional potential surfaces, particularly for the heaviest elements
where the symmetric path seems more probable. There is a very good agreement between the experimental and theoretical 
heights $E_a$ and $E_b$ of the two peaks. The predicted value of the second minimum energy is a little too high. Two reasons
may be advanced to explain this difference. Firstly, the shell energy may be underestimated at this large deformation.
Secondly, the shape sequence may introduce too rapidly the asymmetry, which would 
lead to an undervaluation of the proximity energy. The still sparse
data for the third barrier are correctly reproduced. For the heaviest nuclei the external barrier disappears since the
attractive proximity forces can no more compensate for the repulsive Coulomb forces.    

\newpage
\begin{tabular}{|c|c|c|c|c|c|c|c|c|}
\end{tabular}

Table 1. Comparison between theoretical and experimental \cite{bjor80,blon84,wage91,krasz99} barrier 
characteristics for actinide nuclei.
$E_a$, $E_b$ and $E_c$ are the first, second and third peak heights while $E_{II}$ 
and $E_{III}$ are the energies of the second and third potential minima 
relatively to the ground state energy (in MeV). 

\section{Third barrier}
\begin{figure}[htbp]
\begin{center}
\begin{tabular}{cc}
\includegraphics[height=5.5cm]{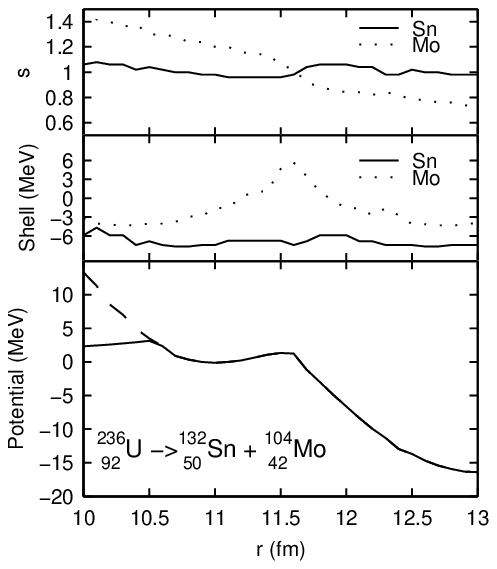} &
\includegraphics[height=5.5cm]{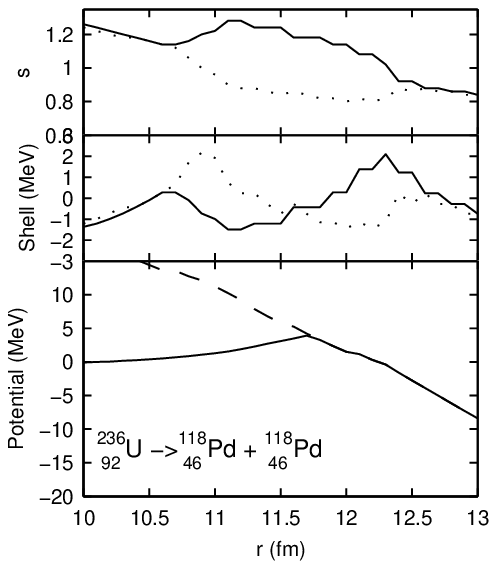}
\end{tabular}
\caption{Fission barriers, shell energies and ratio of the semi-axes of the two ellipsoidal fragments for 
an asymmetric decay channel and the  symmetric one for $^{236}$U. On the lowest part, the fission barrier is
given by the solid line.}
\label{u236symasym}
\end{center}
\end{figure}

The origin of the existence of the third well in the asymmetric decay path is examined now (see Fig. 
\ref{u236symasym}). The dashed line represents the potential for two touching ellipsoids when the one-body
shape is still energetically favoured.  
The second peak (but first on the figure) corresponds to the point where these
 touching ellipsoids begin to give the lowest energy. The 
heaviest fragment is a magic nucleus. It therefore preserves its almost spherical shape. The non magic
fragment was born in an oblate shape ($s \sim 1.4$), due to the small distance between the mass centres at this step.
 When this distance increases, the ratio s decreases, because of the proximity energy which tends to keep close the two
tips of the fragments. Thus, the lightest fragment remaining in contact with the other spherical fragment 
approaches the spherical shape and its shell energy increases to reach a maximum which is at the origin of the 
third peak and which corresponds to two touching different spheres. Before reaching this third peak a third
minimum appears. Its shape is hyperdeformed and asymmetric in agreement with the experimental data 
\cite{krasz99}. Later on, the proximity forces maintain
the two fragments in contact and the shape of the smallest one evolves to prolate shapes ($s<1$) and the shell 
corrections decrease. The third barrier appears only in the asymmetric decay path and for some specific nuclei.
In the symmetric mass exit path, the proximity and Coulomb energies counterbalance the smallest shell effects 
and induce an asymmetric shape, the two fragments remain in contact but one fragment is oblate while the other one is prolate. With 
increasing distance between the mass centres the two nuclei become prolate.   
  
\begin{figure}[htbp]
\begin{center}
\begin{tabular}{cc}
\includegraphics[height=5.5cm]{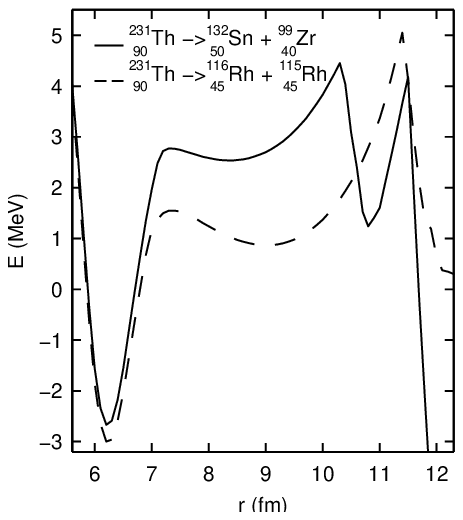} &
\includegraphics[height=5.5cm]{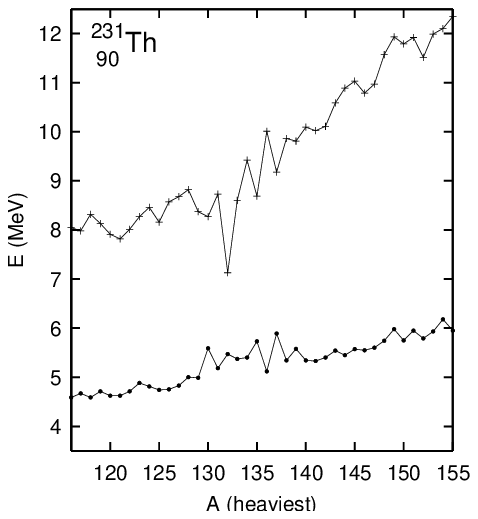} \\
\includegraphics[height=5.5cm]{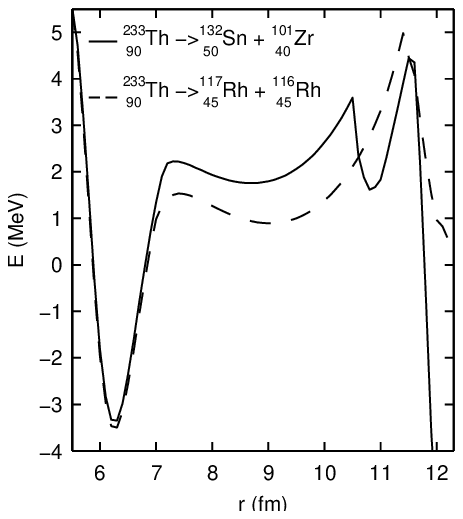} &
\includegraphics[height=5.5cm]{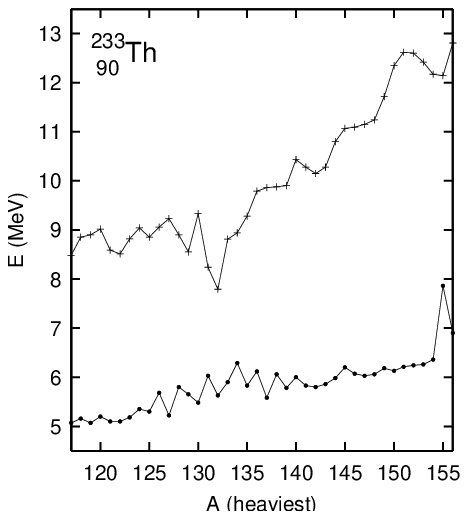} \\
\end{tabular}
\caption{Fission barriers for $^{231,233}$Th and two different decay channels.
The inner and outer fission barrier heights are given, on the right, respectively by the full circles
and crosses.}
\label{th231233}
\end{center}
\end{figure}

\begin{figure}[htbp]
\begin{center}
\begin{tabular}{cc}
\includegraphics[height=5.5cm]{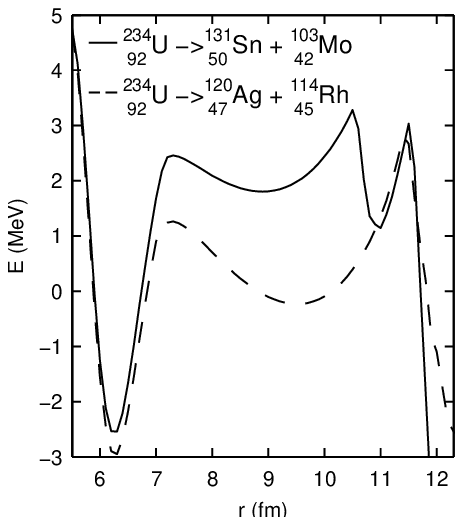} &
\includegraphics[height=5.5cm]{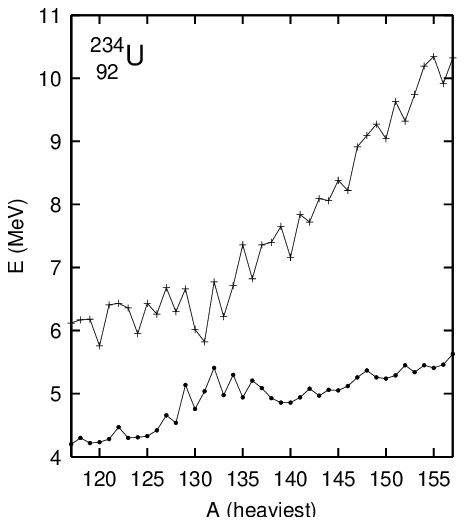} \\
\includegraphics[height=5.5cm]{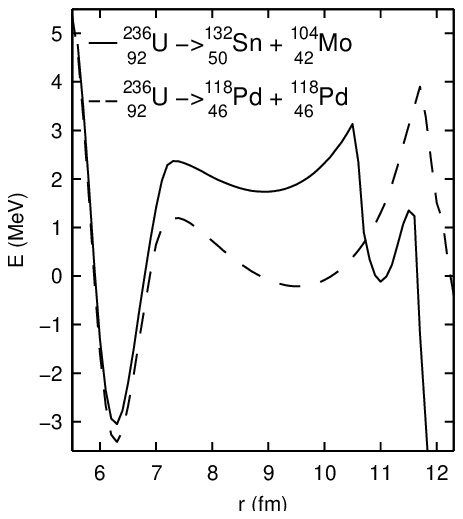} &
\includegraphics[height=5.6cm]{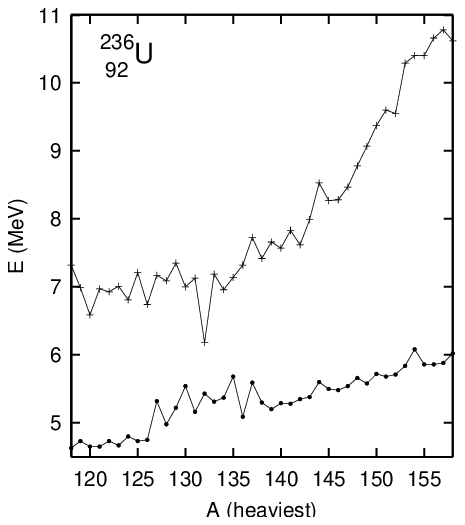} \\
\end{tabular}
\caption{Fission barriers for $^{234,236}$U.
The inner and outer fission barrier heights are given, on the right, respectively by the full circles
and crosses.}
\label{u234236}
\end{center}
\end{figure}

The dependence of the fission barrier heights and profiles on the asymmetry for the $^{231,233}$Th 
and $^{234,236}$U nuclei, 
for which experimental data on the third barrier exist, are given in Fig. \ref{th231233} and \ref{u234236}.
The position of the second peak in the symmetric decay path corresponds to the position of the third peak
in the asymmetric deformation path. Clearly the magicity of some Sn isotopes plays the main role. 

\section{Half-lives}
\begin{figure}[htbp]
\begin{center}
\includegraphics[height=10cm]{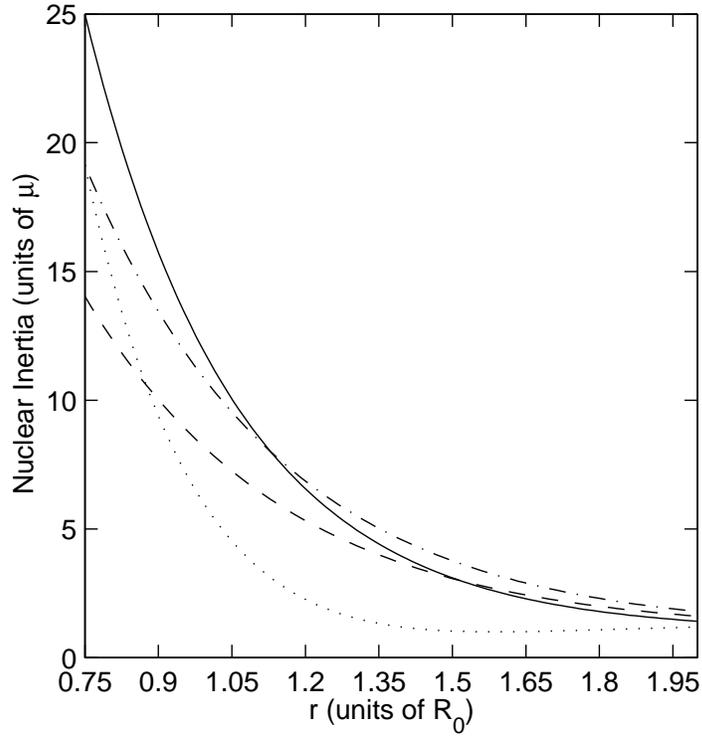}
\caption{Comparison between different selected semi-empirical inertia B(r). 
The dashed line corresponds to the inertia proposed in ref. \cite{rand76} while
the dashed and dotted curve and dotted line give respectively the inertia proposed in ref. \cite{moll89}
 for elongated shapes and for compact and creviced shapes. The values obtained from
 the formula (29) are given by the solid line.}
\label{inertia}
\end{center}
\end{figure}

Within this asymmetric fission model the decay constant is simply 
given by $\lambda=\nu_0P$. The assault frequency $\nu_0$ has been taken as 
$\nu_0=10^{20}\ s^{-1}$.  
The barrier penetrability P is calculated within the action integral
\begin{equation}
P=exp\lbrack -\frac{2}{\hbar}\int_{r_{in}}^{r_{out}}
\sqrt{2B(r)(E(r)-E_{g.s})} dr\rbrack.
\end{equation}    
\newline  
The inertia B(r) is related to the reduced mass by \begin{equation}
B(r)=\mu\lbrack 1+24exp(-3.25(r-R_{sph})/R_0)\rbrack 
\end{equation}  
where $R_{sph}$ is the distance between the mass centres of the future fragments in the initial sphere, 
$R_{sph}/R_0=0.75$ in the symmetric case.\newline 
The inertia depends on the internal structure of the system. A large amount of internal reorganization occuring 
at level crossings raises the inertia. For shapes near the ground state the inertia is expected to be considerably
above the irrotational flow value. For shapes remaining a long time highly deformed the reduced mass is reached
asymptotically. The selected phenomenological inertia B(r) adjusted to reproduce the experimental data is compared 
with other previous semi-empirical inertia in Fig. \ref{inertia}. Its value is slightly higher at the beginning of 
the fission process. It reaches more rapidly the reduced mass value for elongated shapes.\newline  
The partial half-life is finally obtained by $T_{1/2}=\frac{ln2}{\lambda}$.  

The experimental spontaneous fission half-lives and theoretical predictions for the supposed most probable 
exit channels are compared in Table 2. The half-lives corresponding to several paths are given for 
$^{243}$Cm, $^{250}$Cf, $^{256}$Fm and $^{256}$No. Except for the ligthest U isotopes, there is a correct
agreement with the experimental data on 20 orders of magnitude. The half-lives vary regularly for close exit
channels of a same nucleus.

\begin{tabular}{|c|c|c|}
\hline
Reaction & $T_{1/2,exp(s)}$ & $T_{1/2,theo(s)}$     \\
\hline
$^{232}_{92}U \rightarrow \; ^{134}_{52}Te+^{98}_{40}Zr$ & $2.5 \times 10^{21}$ & $3.6 \times 10^{16}$ \\	
\hline
$^{234}_{92}U \rightarrow \; ^{131}_{50}Sn+^{103}_{42}Mo$ & $4.7 \times 10^{23}$ & $8 \times 10^{19}$ \\	
\hline
$^{235}_{92}U \rightarrow \; ^{131}_{50}Sn+^{104}_{42}Mo$ & $3.1 \times 10^{26}$ & $7.7 \times 10^{23}$ \\
\hline
$^{236}_{92}U \rightarrow \; ^{132}_{50}Sn+^{104}_{42}Mo$ & $7.8 \times 10^{23}$ &$1.0 \times 10^{22}$ \\	
\hline
$^{238}_{92}U \rightarrow \; ^{132}_{50}Sn+^{106}_{42}Mo$ & $2.6 \times 10^{23}$ & $5.3 \times 10^{22}$ \\	
\hline
$^{238}_{94}Pu \rightarrow \; ^{130}_{50}Sn+^{108}_{44}Ru$ & $1.5 \times 10^{18}$ & $2.6 \times 10^{19}$ \\
\hline
$^{239}_{94}Pu \rightarrow \; ^{130}_{50}Sn+^{109}_{44}Ru$ & $2.5 \times 10^{23}$ & $4.8 \times 10^{22}$ \\	
\hline
$^{240}_{94}Pu \rightarrow \; ^{130}_{50}Sn+^{110}_{44}Ru$ & $3.7 \times 10^{18}$ & $4.8 \times 10^{19}$ \\
\hline
$^{243}_{95}Am \rightarrow \; ^{133}_{51}Sb+^{110}_{44}Ru$ & $6.3 \times 10^{21}$ & $1.1 \times 10^{23}$ \\	
\hline
$^{243}_{96}Cm \rightarrow \; ^{130}_{50}Sn+^{113}_{46}Pd$ & $1.7 \times 10^{19}$ & $3 \times 10^{21}$ \\	
\hline
$^{243}_{96}Cm \rightarrow \; ^{122}_{48}Cd+^{121}_{48}Cd$ & $1.7 \times 10^{19}$ & $1.6 \times 10^{18}$ \\	
\hline
$^{245}_{96}Cm \rightarrow \; ^{130}_{50}Sn+^{115}_{46}Pd$ & $4.4 \times 10^{19}$ & $3 \times 10^{20}$ \\	
\hline
$^{248}_{96}Cm \rightarrow \; ^{130}_{50}Sn+^{118}_{46}Pd$ & $1.3 \times 10^{14}$ & $7.7 \times 10^{15}$ \\	
\hline
$^{250}_{98}Cf \rightarrow \; ^{125}_{49}In+^{125}_{49}In$ & $5.2 \times 10^{11}$ & $1.9 \times 10^{9}$ \\	
\hline
$^{250}_{98}Cf \rightarrow \; ^{132}_{52}Te+^{118}_{46}Pd$ & $5.2 \times 10^{11}$ & $1.2 \times 10^{10}$ \\	
\hline
$^{250}_{98}Cf \rightarrow \; ^{140}_{55}Cs+^{110}_{43}Tc$ & $5.2 \times 10^{11}$ & $4.9 \times 10^{11}$ \\	
\hline
$^{255}_{99}Es \rightarrow \; ^{128}_{50}Sn+^{127}_{49}In$ & $8.4 \times 10^{10}$ & $8 \times 10^{9}$ \\	
\hline
$^{256}_{100}Fm \rightarrow \; ^{128}_{50}Sn+^{128}_{50}Sn$ & $ 1.0 \times 10^{4}$ & $ 45$ \\	
\hline
$^{256}_{100}Fm \rightarrow \; ^{121}_{47}Ag+^{135}_{53}I$ & $ 1.0 \times 10^{4}$ &  82 \\	
\hline
$^{256}_{102}No \rightarrow \; ^{128}_{51}Sb+^{128}_{51}Sb$ & $ 110 $& $0.9 \times 10^{-2}$ \\	
\hline
$^{256}_{102}No \rightarrow \; ^{116}_{46}Pd+^{140}_{56}Ba$ & $ 110 $& $0.3 \times 10^{-1}$ \\	
\hline
\end{tabular}
\newline
\newline
Table 2. Comparison between experimental \cite{wage91} and theoretical \\ spontaneous fission half-lives 
of actinide nuclei.

\section{Summary and conclusion}
Double and triple-humped fission barriers appear for the actinide elements in the quasi-molecular shape path 
within a macroscopic-microscopic deformation energy
 derived from a generalized liquid drop model and analytic expressions for the shell and pairing energies. 
The second peak corresponds to the transition from one-body shapes to two touching ellipsoids. 
The third barrier appears only in the asymmetric decay path and for some specific nuclei. 
Then, the heaviest fragment is almost a magic nucleus and it preserves its shape close to the sphere. 
The other fragment evolves from an oblate ellipsoid to a prolate one and the third peak corresponds to 
the maximum of the shell effects in the non magic fragment and, consequently, to two touching different 
spheres. The barrier heights agree
with the experimental results for the double and triple-humped fission barriers, the energy of the second minimum 
being a little too high. The predicted half-lives follow the experimental data trend.

\end{document}